\documentclass{ws-procs9x6square}

\begin{document}


\title{Quantum Hall Physics in String Theory\footnote{\uppercase{T}o 
appear in the proceedings
of the 3rd \uppercase{I}nternational \uppercase{S}ymposium on 
\uppercase{Q}uantum \uppercase{T}heory and \uppercase{S}ymmetries 
(\uppercase{QTS}3),
\uppercase{S}ept. 10 - 14, 2003.}}

\author{Oren Bergman}
\address{Department of Physics\\
Technion, Israel Institute of Technology\\
Haifa 32000, Israel}  



\maketitle

\abstracts{In certain backgrounds string theory exhibits quantum Hall-like behavior. 
These backgrounds provide an explicit realization of the effective non-commutative 
gauge theory description of the fractional quantum Hall effect (FQHE), and of the 
corresponding large $N$ matrix model. I review results on the string theory realization of
the two-dimensional fractional quantum Hall fluid (FQHF), and describe new results on the
stringy description of higher-dimensional analogs.}


\section{Introduction}

Two  of the most exciting developments in physics in the last 20 years
have been the (experimental) discovery of the fractional quantum Hall 
effect (FQHE) \cite{Tsui:1982yy}, and 
the (theoretical) discovery of superstring theory \cite{Green:sg}.
The two disciplines, and even 
more so their practitioners, could not have less to do with each other however.
Or so it seemed. In this lecture I will describe 
configurations in superstring theory
which exhibit the fractional quantum
Hall effect,
thereby establishing a connection 
between these two fields of physics. 
This connection suggests
a new direction for string theory, namely as an
{\em effective} theory of the FQHE. In no way do I mean to imply that string theory
should replace the Coulomb Hamiltonian as the microscopic physics 
underlying the FQHE. I simply propose that string theory, in the particular configurations
described below, may present an improvement over other theories as an effective
long-wavelength description of the FQHE. 
String theory in fact implies the effective gauge theory
of the FQHE. In particular, I will show that the string
theory picture provides a derivation of Susskind's recent proposal that the ground
state of the FQHF is described by a non-commutative Chern-Simons
(NCCS) gauge theory \cite{Susskind:2001fb}. 
This, together with the direct derivation of the quantization of
the inverse filling fraction, and the fractionally charged quasiparticles,
serves as evidence for my proposal.

Most of this lecture follows work with John Brodie and Yuji Okawa
\cite{Bergman:2001qg}, and a continuing collaboration with John Brodie
\cite{Bergman_Brodie}.

\section{Branes and the 2d FQHE}

\subsection{The D0-brane quantum Hall fluid}

\begin{figure}[ht]
\centerline{\epsfxsize=4in\epsfbox{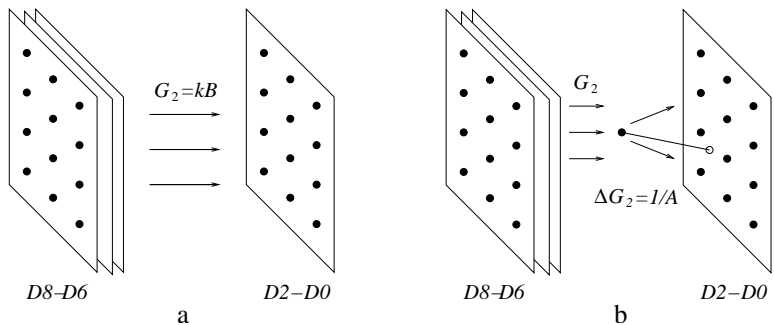}}   
\caption{(a) The D0-brane quantum Hall fluid: RR charged D0-branes in two dimensions 
with a uniform magnetic RR field $G_2$. (b) A quasiparticle excitation.}
\label{D2D8}
\end{figure}

The configuration for the two-dimensional FQHF consists of a D2-brane in
the background of an even number $2k$ of D8-branes, and a $B$-field 
$b$ along the D2-brane (Figure~\ref{D2D8}a). 
The D8-brane background implies that the low-energy effective spacetime theory
is massive IIA supergravity \cite{Romans}, where
the cosmological constant is given by the RR 0-form field strength 
$G_0 = k$ \cite{Polchinski_RR}.
The $B$-field has two effects: it induces D6-brane charge in the 
D8-branes, and D0-brane charge in the D2-brane. This is a consequence of the
Chern-Simons term in the D$p$-brane world-volume theory,
\begin{equation}
  S_{CS} = \int C\wedge e^{F + B} \;,
\label{CS}
\end{equation}
where $C$ represents a formal sum of 
all the odd-degree RR potentials.
The induced D0-brane and D6-brane charge densities are therefore given by
$\rho_0 = b$ and $\rho_6 = 2kb$, respectively.
The D6-branes are a source of a uniform magnetic RR field perpendicular to
the D2-brane, $G_2 = dC_1 = kb$.
We therefore end up with a system of charged particles, the D0-branes,
moving in two dimensions, {\em i.e.} in the D2-brane, in a uniform background magnetic
field. The Landau level filling fraction is given by
\begin{equation}
 \nu = {\rho_0\over G_2} = {1\over k} \;.
\label{filling_fraction}
\end{equation}
Furthermore, it can be shown that the two-dimensional statistics of the D0-branes
are fermionic for odd $k$ and bosonic for even $k$.
This configuration therefore describes the primary Laughlin
state of the FQHF at filling fraction $1/k$ \cite{Laughlin},
with the D0-branes playing the role of the electrons.

\subsection{The D2-brane effective gauge theory}

The above system can also be studied from the point of view of the D2-brane
world-volume. The low energy dynamics of the D2-brane is described by the
three-dimensional world-volume gauge theory,
\begin{equation} 
S_{D2} = S_{DBI} + S_{CS} 
 + {k\over 4\pi}\int A\wedge F\;,
\label{D2action}
\end{equation}
where $S_{DBI}$ is the Dirac-Born-Infeld action,
and the additional Chern-Simons term is a result of integrating out massive
fermions coming from the D2-D8 strings. 
In order to decouple the world-volume theory from the bulk we need
to scale the various parameters and fields in a particular way
as we take the low-energy limit $\alpha'\rightarrow 0$. There
is actually some freedom here, and we choose the following scaling
behavior: 
\begin{equation}
 b \sim 1 \;,\;
 C_n \sim (\alpha')^{n/2} \;,\;
 g_s \sim  (\alpha')^{3/2+\epsilon} \;,\;
 G_{ij}\sim (\alpha')^{2+\delta} \;\; (i,j=1,2) \;,
\label{scaling}
\end{equation}
where $\delta>\epsilon>0$. 
Apart from the contributions of the background D2-brane
and D0-brane charges, the action reduces to
\begin{equation}
 S_{D2} = {k\over 4\pi} \int A\wedge F  +
             {1\over 2\pi}\int C_1\wedge F\;.
\label{CS_action}
\end{equation}
What we have obtained in (\ref{CS_action}) is known as 
the {\em effective hydrodynamic gauge theory}
of the quantum Hall fluid \cite{Bahcall:an,Fradkin:2002qw}. The world-volume
gauge field $A$ plays the role of the {\em hydrodynamic gauge field},
which describes the fluctuations of the fluid, and the RR field $C_1$
plays the role of the electromagnetic gauge field.
The second term corresponds simply to the coupling of the charge
to the external electromagnetic field.
The special properties of the fluid are encoded in the first term,
which is an ordinary Abelian CS action.

The scaling behavior described above is a generalization of the Seiberg-Witten
limit \cite{Seiberg:1999vs}, which corresponds to $\delta = \epsilon =0$.
In this case the world-volume gauge theory is most naturally
expressed as a {\em non-commutative gauge theory},
in terms of a new gauge field $\widehat{A}$, 
and a new kind of product called the Moyal star product.
In our case the CS action is re-expressed as a non-commutative
CS action,
\begin{equation}
 \hat{S}_{NCCS} = {k\over 4\pi}\int
 \left(\hat{A}\star d\hat{A} 
+ {2i\over 3}\hat{A}\star\hat{A}\star\hat{A}\right)\;.
\end{equation}
This reproduces Susskind's improved proposal that, in order
to capture the graininess of the quantum Hall fluid, one needs
to elevate the effective hydrodynamic gauge theory to a 
non-commutative gauge theory \cite{Susskind:2001fb}.

\subsection{The quasihole/quasiparticle excitations}

The lowest lying excitations of the fractional quantum Hall fluid
are fractionally charged states which carry a unit of magnetic flux.
These are known as quasiparticles or quasiholes, depending on the sign
of the charge and flux.
We create such an excitation in the string theory picture by taking
an additional D6-brane across the D2-brane (Figure~\ref{D2D8}b). 
The D6-brane is completely orthogonal to the D2-brane,
in other words the two branes are ``linked'' in the sense of \cite{Hanany:1996ie}. 
This move has two physical 
consequences: it changes the $G_2$ flux by one unit, and it creates a string
between the D6-brane and the D2-brane.
The end of the string actually carries a fractional D0-brane charge $1/k$ in the 
D2-brane.
This can be shown by extending Strominger's charge conservation argument
\cite{Strominger} to D-branes in massive IIA supergravity \cite{Bergman_Brodie}.
This means that the D0-brane density is now
\begin{equation}
 \rho_0(x) = \left( B + {1\over kA}\right)
    - {1\over k}\delta^{(2)}(x) \;,
\end{equation}
where $A$ is the area of the D2-brane.
The shift in the uniform component of the density is due to the change in $G_2$, and
this ensures that the filling fraction of the fluid remains $1/k$.
The localized component is the fractionally charged quasiparticle 
(or rather quasihole in this case).

\subsection{The matrix model}

The role of the electrons in our quantum Hall fluid is played
by D0-branes. We can therefore describe the fluid alternatively
in terms of a large $N$ D0-brane matrix model, which in 
the scaling limit (\ref{scaling}) reduces to
\begin{eqnarray}
 S_{D0} = \int dt\, \mbox{Tr} \left[
 G_2\, X^1 D_0 X^2  - kA_0
+ i\psi^\dagger_a D_0 \psi^{\phantom{\dagger}}_a\right]\;.
\end{eqnarray}
The first term is the Lorentz interaction of the D0-branes with the
RR 2-form field strength, the second term is a one-dimensional CS
term which comes from the coupling of the D0-branes to the RR 0-form
field strength, and the last term corresponds to the D0-D8
fermions. This is precisely the matrix model proposed by 
Polychronakos as an alternative formulation of the NCCS theory 
of the FQHF \cite{Polychronakos:2001mi}.
The allowed D0-brane configurations are solutions of the constraint
\begin{equation}
 G_2[X^1,X^2] + i\psi^\dagger\psi
  =ik{\bf 1}\;.
\end{equation}
A D2-brane is a large $N$ solution for which
$\mbox{Tr}[X^1,X^2]=iA$.
In particular, the ground state of the FQHF is given by
\begin{equation}
[X^1,X^2]={iA\over N}{\bf 1} \;,\;
\psi^\dagger\psi=0\;,
\end{equation}
and a FQHF with a quasihole excitation is given by
\begin{equation}
[X^1,X^2]={iA\over N + {1\over k}}\left({\bf 1}
 +{1\over k}|0\rangle\langle 0|\right)\;,\;
\psi^\dagger\psi = - |0\rangle\langle 0| \;.
\end{equation}

\section{Branes and higher-dimensional FQHFs}

Higher dimensional analogs of the FQHE have recently been proposed.
These include non-Abelian particles moving on $S^4$ in a background
$SU(2)$ instanton gauge field \cite{Zhang:2001xs}, as well as ordinary
charged particles on $CP^3$ \cite{Bernevig:2002eq}, and more generally
$CP^n$ \cite{Karabali:2002im}, in a background $U(1)$ magnetic field.
In the large volume limit the latter system becomes 
a planar $2n$-dimensional quantum Hall fluid with a uniform magnetic 
field in each of the $n$ independent planes. The main properties
of these generalized QHFs are:
\begin{enumerate}
\item Quantized inverse filling fraction $\nu = 1/k^n$, $k$ odd.
\item Fractionally charged quasiholes/quasiparticles, charge $\pm \nu$.
\item Brane excitations with fractional statistics.
\end{enumerate}

\begin{figure}[ht]
\centerline{\epsfxsize=4in\epsfbox{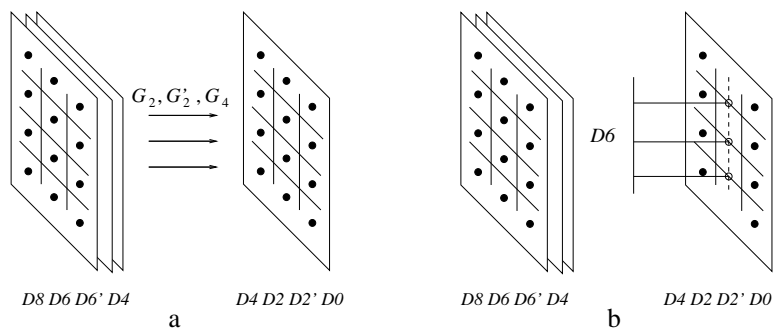}}   
\caption{(a) The 4-dimensional D0-brane quantum Hall fluid. The vertical and horizontal
lines represent the two kinds of D2-branes (D6-branes) in the D4-brane (D8-branes),
and the dots represent the D0-branes (D4-branes) in the the D4-brane (D8-branes).
(b) A quasimembrane excitation of the 4-dimensional FQHF.}
\label{D4D8}
\end{figure}

String theory provides a natural generalization of the planar QHF 
to four and six dimensions, by replacing the D2-brane with a D4
or D6-brane, respectively. Let me focus on the four-dimensional case,
the generalization to six dimensions should be straightforward.
The configuration consists of a D4-brane in the background
of $2k$ D8-branes and {\em two} non-trivial components of the $B$-field,
$b$, $b'$, along the two planes of the D4-brane. This will induce both D2-branes
and D0-branes in the D4-brane, and D6-branes and D4-branes in the D8-branes
(Figure~\ref{D4D8}a). 
The 4-dimensional analog of the Landau level filling fraction is given by
the ratio of the D0-brane density and the product of the two RR magnetic fields
due to the D6-branes,
\begin{equation}
\nu = {\rho_0\over G_2 G'_2} = {bb'\over (kb)(kb')}
= {1\over k^2} \;,
\end{equation}
which agrees with property 1 above.

This system can also be described in terms of the low-energy effective
gauge theory on the D4-brane world-volume. Adapting the scaling limit
(\ref{scaling}) to the D4-brane gives
\begin{equation}
 S_{D4} = {k\over 24\pi^2} \int A\wedge F^2  +
             {1\over (2\pi)^2}\int C_1\wedge F^2
   + {1\over 2\pi} \int C_3\wedge F \;.
\label{CS_action_D4}
\end{equation}
This is our proposal for the effective gauge theory of the four-dimensional
quantum Hall fluid at filling fraction $1/k^2$.
As in the two-dimensional case, $A$ corresponds to the
hydrodynamic gauge field, and $C_1$ to the electromagnetic gauge field.
The new ingredient here is the 3-form gauge field $C_3$,
whose field strength $G_4$ was also part of the background.
This field did not appear in the original proposals for the 
higher-dimensional QHFs, but it is a very natural extension.

\subsection{Quasimembranes}

The 3-form couples naturally to 
the membrane excitations of the 4-dimensional QHF \cite{Bernevig:2002eq}.
Let me briefly describe how these membranes appear in the stringy construction.
Consider a process similar to the one discussed in the context of the quasiparticles
in the two-dimensional case, namely a D6-brane moving across the D4-brane
(Figure~\ref{D4D8}b).
The D6-brane is ``linked'' with the horizontal D2-branes in the D4-brane,
so strings are created between the D6-brane and these D2-branes. One ends up
with a 2-dimensional array of strings, each of which carries a $1/k$ D0-brane
charge in the D4-brane. This array forms a membrane-like object in the D4-brane.
It turns out that this object also carries $1/k$ 
charge under $C_3$. It therefore corresponds
to a fractional D2-brane, in the same sense that the end of a single
string corresponds to a fractional D0-brane. We therefore call it
a {\em quasimembrane}. Quasimembranes come in two varieties, depending
on whether the D6-brane is linked with the horizontal or vertical D2-branes.

\section{Where do we go from here?}

This program can be extended in a number of directions.
We have considered planar QHFs of infinite extent corresponding
to infinite branes.
It would be interesting to construct other geometries, as well
as finite systems. For example, a QHF on a strip should correspond
to a D2-brane suspended between two NS5-branes. Such a system
should exhibit massless edge excitations, in addition to the massive
quasiparticles.
We would also like to understand how the hierarchy of filling
fractions arises in the stringy picture. There are a couple of
different phenomenological models of the hierarchy 
\cite{Haldane:1983xm,Jain:1989tx}, and the question is which, if any, is 
realized by the stringy QHF.

Brane realizations of gauge theories in string theory have led
to many new insights into the dynamics of gauge theories,
especially through the large $N$ gauge/gravity duality.
The FQHE can also described by an (effective) gauge theory -
one which we were able to realize using branes in string theory.
Is there a FQHE/gravity duality? To answer this question one must
first find the supergravity solution corresponding to the brane
configuration I described. This has not been done yet.
The real test of the usefulness of the string theory/QHE connection is 
whether string theory teaches us something {\em new} 
about the fractional quantum Hall effect.
This is yet to be determined.



\section*{Acknowledgments}

This work was supported in part by the Israel Science Foundation
under grant no. 101/01-1.







\end{document}